\documentclass[a4paper,fleqn,usenatbib]{mnras}
\usepackage[T1]{fontenc}
\usepackage{ae,aecompl}

\usepackage{graphicx}	
\usepackage{amssymb}	

\title[V4743 Sgr, a magnetic nova?]{V4743 Sgr, a magnetic nova?\footnote{
based on observations made with the Southern African Large Telescope (SALT).}
}
\author[P. Zemko, M. Orio, K. Mukai, A. Bianchini, S. Ciroi and V. Cracco]{P. Zemko$^{1}$\thanks{E-mail:
polina.zemko@studenti.unipd.it},
M. Orio$^{2,3}$, 
K. Mukai$^{4,5}$,
A. Bianchini$^{1,2}$,
S. Ciroi$^{1}$
and V. Cracco$^{1}$
\\
$^{1}$Department of Physics and Astronomy, Universit\`a di Padova, vicolo dell' 
Osservatorio 3, I-35122 Padova, Italy\\
$^{2}$INAF - Osservatorio di Padova, vicolo dell' Osservatorio 5, I-35122 Padova, Italy\\
$^{3}$Department of Astronomy, University of Wisconsin, 475 N. Charter Str., 
Madison, WI 53704, USA\\
$^{4}$CRESST and X-ray Astrophysics Laboratory, NASA Goddard Space Flight Center, 
Greenbelt, MD 20771, USA\\
$^{5}$Department of Physics, University of Maryland, Baltimore County, 1000 
Hilltop Circle, Baltimore, MD 21250, USA
}

\date{}

\pubyear{2002}

\begin{document}

\label{firstpage}
\pagerange{\pageref{firstpage}--\pageref{lastpage}}
\maketitle
\begin{abstract}
Two {\sl XMM Newton} observations of Nova V4743 Sgr (Nova Sgr 2002) were performed shortly after it
returned to quiescence, 2 and 3.5 years after the explosion. The X-ray light curves 
revealed a modulation with a frequency of $\simeq$0.75 mHz, indicating that V4743 Sgr is 
most probably an intermediate polar (IP). The X-ray spectra have characteristics in common
with known IPs, with a hard thermal plasma component that can be fitted only
assuming a partially covering absorber. In 2004 the X-ray spectrum had also a supersoft 
blackbody-like component, whose temperature was close to that of the white dwarf (WD) in 
the supersoft X-ray phase following the outburst, but with flux by at least two orders of magnitude
lower. In quiescent IPs, a soft X-ray flux component
originates at times in the polar regions irradiated by an accretion column, but the supersoft 
component of V4743 Sgr disappeared in 2006, indicating a possible origin different from accretion. 
We suggest that it may have been due to an atmospheric temperature gradient on the WD surface,
 or to continuing localized thermonuclear
burning at the bottom of the envelope, before complete turn-off.
An optical spectrum obtained with SALT 11.5 years after
the outburst showed a
prominent He II $\lambda$4686 line and the Bowen blend, which reveal a very hot 
region, but with peak temperature shifted to the ultraviolet (UV) range. 
V4743 Sgr is the third post-outburst nova and IP candidate 
showing a low-luminosity supersoft component in the X-ray flux a few years after the outburst.

\end{abstract}


\section{Introduction}

 Nova V4743 Sgr was discovered by \citet{has02V4743Sgr} in outburst in 2002 September 20 
 close to the 5th magnitude. \citet{kat02V4743Sgr} classified this object as an 
  Fe II-class nova and found that the full width at half maximum (FWHM) of the H$\alpha$ 
emission was 2400\,km\,s$^{-1}$. 
t$_2$ and t$_3$ times (times needed to fade by 2 and 3 
magnitudes from maximum) for V4743 Sgr are 6 and 12 days, respectively, typical of a very fast nova \citep{str10novae}.
Even if the optical light curve decay was fast, the X-ray light curve developed 
relatively slowly -- the nuclear-burning phase lasted at least for 1.5 years after the outburst \citep{rau10V4743Sgr}, 
unlike, for instance, the recurrent novae (RNe), which seem to burn the remaining hydrogen
very rapidly after the outburst. 

V4743 Sgr was the first nova regularly monitored with X-ray gratings in outburst 
\citep{sta03V4743Sgr, nes03V4743Sgr, lei06V474Sgr, rau10V4743Sgr}.
Five aditional X-ray grating observations were obtained between 2002 March and 2004 April with 
{\sl Chandra} and {\sl XMM Newton} and four of them coincided with the supersoft X-ray source 
(SSS) phase of the nova, when the ejecta became transparent to X-rays from the central source. During
the SSS phase the nova was the brightest supersoft X-ray source in the sky 
and had a continuous spectrum with deep absorption features of O, Ni, and C \citep{nes03V4743Sgr}.
\citet{rau10V4743Sgr} analysed the grating spectra of V4743 Sgr using a Non-Local
 Thermal Equilibrium (NLTE) Atmosphere Model \citep{rau03NLTE} and found that
the nova reached its highest effective temperature of 740 000 K around 2003 April
 and remained hot for at least 5 months. With such a peak temperature the WD is 
 1.1 -- 1.2 M$_{\sun}$. \citet{van12V4743Sgr} found a lower value of 
 the effective temperature -- 550 000 K, applying a wind-type expanding NLTE model \citep{van10model}. 
The difference in the derived effective temperature is not only due to the applied models
but also to the higher value of N(H) inferred by \citet{van12V4743Sgr}. Moreover, the model of \citet{van12V4743Sgr}
was developed with solar abundances, which are, however, not suitable for a WD atmosphere.

 In addition to offering a view of the WD atmosphere and its composition, the long exposure
 times in the X-ray grating spectra allowed
 detection of intense X-ray variability. The nova was in fact very variable, both aperiodically
 and periodically. \citet{nes03V4743Sgr} detected large-amplitude oscillations with a 
 period of 1325 s (22 minutes) in the first half of their 25 ks {\sl Chandra} exposure but
 soon after the count rate suddenly dropped to a very low value until the end of the exposure,
and only emission lines were observed.
 Such a low state has never been observed in any of subsequent observations of V4743 Sgr
 and was probably caused by a temporary obscuration of the central X-ray continuum source by 
 clumpy ejecta \citep{nes13obs}.
 \citet{lei06V474Sgr} analysed the periodic variability of the X-ray flux in the first two
  X-ray observations during the outburst, spaced two weeks apart and detected a combination of 
  oscillations represented by a number of discrete frequencies lower than 1.7 mHz. At least
 five of these frequencies, including the one at 0.75 mHz, were present in both observations.
  The authors proposed that the 0.75 mHz frequency and its first harmonic in the power spectrum
 were related to the WD spin period while the other oscillations were due to non-radial 
 pulsations of the WD. 
\citet{dob10V4743Sgr} extended the study of the X-ray variability of V4743 Sgr, including 
three more observations of the nova in outburst, and the two quiescent exposures, proposed by 
us and discussed in this paper. These authors
found that the 0.75 mHz feature in the power spectrum was due to different frequencies in
outburst, but eventually in the quiescent observations only one frequency remained, which 
they attributed to the spin period of a magnetized WD in an intermediate polar (IP), following
 also a suggestion of \citet{lei06V474Sgr}. 
Optical observations also strongly support the IP scenario. \citet{kan06V4743Sgr}, \citet{ric05V4743Sgr}
and \citet{wag03V4743Sgr} presented measurements of the orbital period ($\sim$ 6.7 h), and detected
 a much shorter period of $\sim$24 minutes, which seems to be the beat of the orbital period and 
 the one observed in the X-rays. 
 
From their infrared observations \citet{nie03v4743sgr} roughly estimated the distance to 
V4743 Sgr as 1200$\pm$300 pc, however, the authors pointed out that this value should be
 taken with caution because of the uncertainties in the estimates of the interstellar extinction
towards the nova. The authors also stressed that the maximum possible distance is 6 kpc.
\citet{van07V743Sgr} derived a distance of 3.9$\pm$0.3 kpc using the maximum-magnitude 
rate-of-decay (MMRD) relationship \citep{val95mmrd}.
 
In this paper we present the {\sl XMM Newton} observations, proposed by us (PI Orio) and 
performed in quiescence 2 and 3.5 years after the nova explosion and the optical spectra of 
V4743 Sgr, obtained with the Southern African Large Telescope (SALT) 12 years after the 
outburst, revealing the evolution of the supersoft X-ray component and the expanding nova shell.

\section{X-ray observations and data analysis}
V4743 Sgr was observed with all the instruments onboard {\sl XMM Newton} 
 on September 30 2004 and with the European Photon Imaging Camera (EPIC),
and the Reflection Grating Spectrometer (RGS 1 and RGS 2) on March 28 2006 (742 and 1286 days 
after the nova explosion, respectively). Here we focus mainly on the data
of the EPIC Metal Oxide Semi-conductor (MOS) CCD arrays: MOS 1 and MOS 2. The RGS data had
a very low signal-to-noise ratio and only the 2004 observation allowed to marginally 
detect several emission lines. The data of the EPIC pn camera were not suitable for the 
analysis, since the source was on a chip gap. The X-ray spectra were fitted with XSPEC v.12.8.2.
 The dates and exposure times of both X-ray and optical observations are 
 presented in Table 1.

 \begin{table*}
\clearpage
 \centering
 \begin{minipage}{140mm}
  \caption{Observational log of the {\sl XMM Newton} and {\sl SALT} observations of V4743 Sgr.}
  \begin{tabular}{llcc}
\hline
Date and time & Instrument & Exposure (s) & Count rate (cnts s$^{-1}$)/Mag.$^1$ \\
\hline
\hline
2004-09-30 18:28:25 &{\sl XMM Newton} MOS1 	 & 22163 & $0.1013\pm0.0022$\\
2004-09-30 18:28:23 &{\sl XMM Newton} MOS2 	 & 22168 & $0.1031\pm0.0022$\\  
2004-09-30 18:27:28	&{\sl XMM Newton} RGS1 	 & 21295 & $0.0058\pm0.0022$\\  
2004-09-30 18:27:36 &{\sl XMM Newton} RGS2 	 & 21295 & $0.0063\pm0.0024$\\  
2004-09-30 18:36:34 &{\sl XMM Newton} OM U 	 & 4000  & $14.4019\pm0.0015$\\  
2004-09-30 19:48:21 &{\sl XMM Newton} OM B 	 & 4000  & $15.5172\pm0.0027$\\  
2004-09-30 21:00:09 &{\sl XMM Newton} OM UVW1& 3998  & $14.0620\pm0.0015$\\  
2004-09-30 22:11:55 &{\sl XMM Newton} OM UVM2& 4181  & $14.156\pm0.013$\\  
2004-09-30 23:26:56 &{\sl XMM Newton} OM UVW2& 4398  & $14.265\pm0.020$\\  
\hline
2006-03-28 15:28:18 &{\sl XMM Newton} MOS1 	 & 34164 & $0.0597\pm0.0017$\\  
2006-03-28 15:28:18 &{\sl XMM Newton} MOS2   & 34168 & $0.0692\pm0.0017$\\  
2006-03-28 15:27:25 &{\sl XMM Newton} RGS1 	 & 34418 & $0.0009\pm0.0012$\\  
2006-03-28 15:27:29 &{\sl XMM Newton} RGS2 	 & 34410 & $0.0019\pm0.0015$\\  
\hline
2014-03-21 02:54:48	&{\sl SALT} RSS$^2$		  & 1000  & \\  
2014-03-21 03:11:48 &{\sl SALT} RSS		  & 1000  & \\  
\hline
\multicolumn{4}{p{.9\textwidth}}{{\bf Notes.}{\footnotesize $^1$ The mean count rate during
 the exposure for the X-ray observations and the mean magnitude for the {\sl XMM Newton} OM
optical monitor (OM) observations.
The effective wavelengths of the {\sl XMM Newton} OM filters are:
U --- 344 nm, B --- 450 nm, UVW1 --- 291 nm, UVM2 --- 231 nm, UVW2 --- 212 nm.
$^2$Robert Stobie Spectrograph}}\\
\end{tabular}
\end{minipage}
\label{tab:obs}
\end{table*}

\subsection{ Spectral analysis}\label{sseq:xrayspec}
\begin{figure*}
\centering
\includegraphics[width=300pt]{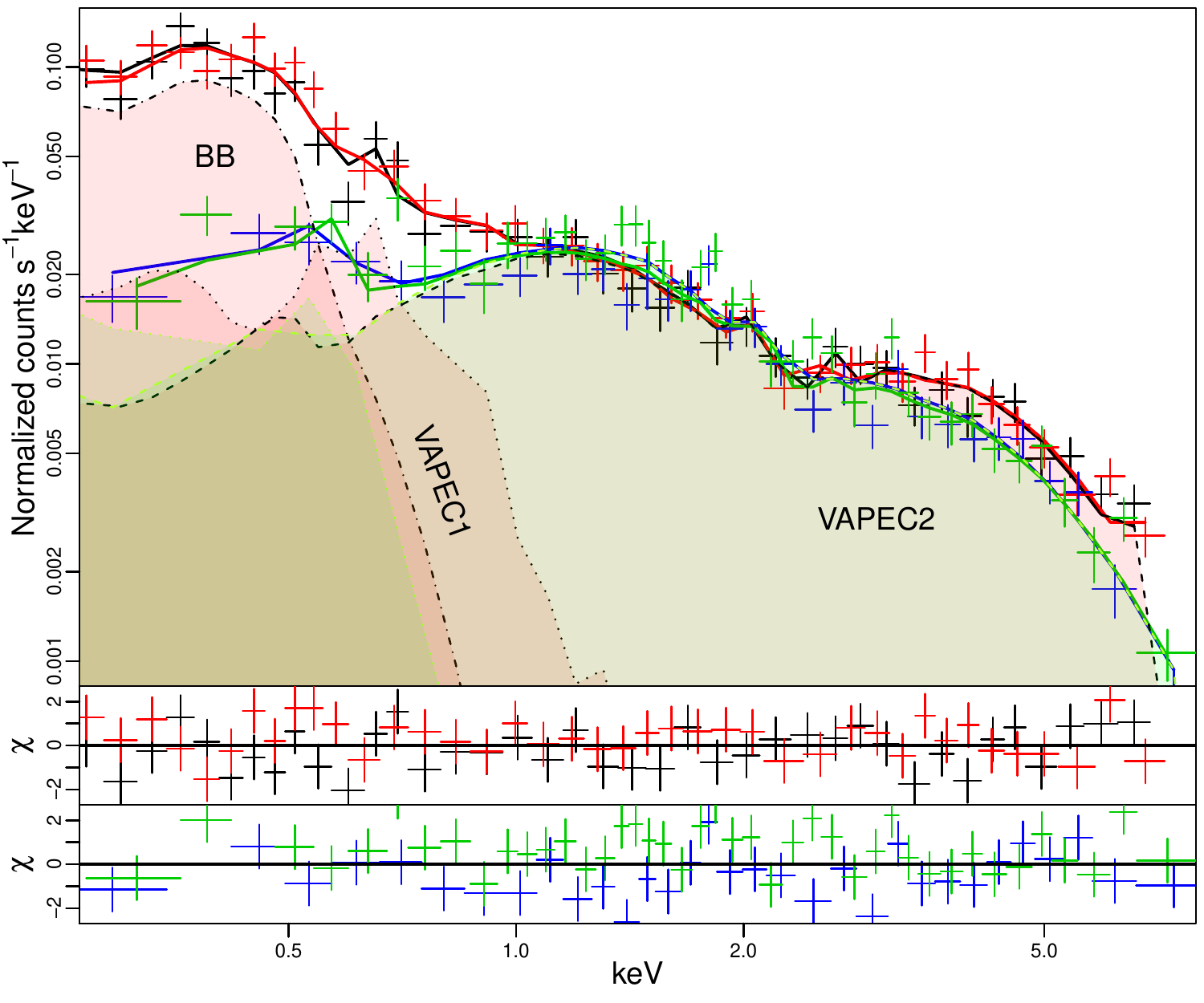}
\caption{Top panel: XMM {\sl Newton} spectra of V4743 Sgr. The 2004 data are plotted in black (MOS 1 data)
 and in red (MOS 2) and the 2006 data in blue (MOS 1) and in green (MOS 2). 
The solid lines represent the fit with wabs$\times$(bb+pcfabs$\times$(vapec+vapec))
model, where the parameters of the interstellar absorption (wabs) and the plasma abundances of the 
2004 and 2006 spectra were tight. The normalization of the blackbody component was set to 
zero in the model of the 2006 spectra. The components of the models are plotted with the dotted (low-temperature
vapec components), dashed (high-temperature vapec components) and the dash-dotted lines (the blackbody component).
The model components of the 2004 data were filled with pink and that of the 2006 model -- with green. Middle panel: residuals 
of the 2004 MOS 1 and MOS 2 data fit. Bottom panel: residuals of the 2006 MOS 1 and MOS 2 data fit.}
\label{fig:spectra}
\end{figure*}

 The background subtracted 0.2 -- 10.0 keV spectra of V4743 Sgr together with the best-fitting
models are presented in Figure~\ref{fig:spectra}. The 2004 data are plotted in black (MOS 1)
 and red (MOS 2) and the 2006 data in blue (MOS 1) and green (MOS 2). 
We first analysed the 2004 spectrum because of the higher count rate and found that the
best-fitting model consists of a blackbody and two thermal plasma components affected by 
a partially covering absorber ({\it PCFABS} model in the XSPEC). Without the complex absorption 
it was impossible to fit the hard part of the spectra even increasing the
temperature. We used the {\it VAPEC} model of thermal plasma emission in XSPEC,
since it allows to constrain abundances of different elements and novae have a highly non-solar
composition.

 In order to compare our results with the outburst 
{\sl XMM Newton} and {\sl Chandra} X-ray spectra of V4743 Sgr studied in detail by 
\citet{rau10V4743Sgr} and \citet{van12V4743Sgr} we attempted
a fit of the supersoft component in the 2004 spectrum with the atmospheric model
 of \citet{rau10V4743Sgr}. The 
atmospheric model that we used had the same abundances and $\log{g}$ as 
model B in \citet{rau10V4743Sgr}. 

The next step was to fit the 2006 data. From fig.~\ref{fig:spectra} we see that major changes during one and a hlaf year
 between observations were in the 
softest region of the spectrum --- the blackbody-like component was no longer measured in the
2006 observation, while the part of the spectrum above 1 keV was almost the same. The data 
quality did not allow us to fit the 2006 spectrum independently, so we tried a simultaneous
fit of the 2004 and 2006 datasets applying the same model and setting the 
normalization of the blackbody component to zero. We also assumed that the interstellar absorption
and the element abundances of the plasma are the same in both observations but let other parameters
to vary freely. Although the variable opacity of the nova shell may 
contribute to the value of the interstellar absorption, we did not expect the absorption to increase,
obscuring the soft emission in 2006. 

The best fitting parameters of our models are
summarised in Table 2. All the fits required an increased abundances of Si and S, but the 
spectra quality did not allow us to constrain their values. We also did not detect the 6.4 keV 
Fe K$\alpha$ reflection line at a significant level.

\citet{rau10V4743Sgr} fitted the spectrum during the constant bolometric luminosity phase with
T$_{eff}\simeq$ 700 000 K and found only evidence of moderate cooling with a temperature of 660 000 K
in February of 2004, but the temperature may have been constant within the errors.
Eight months later, in the first observation we discuss here, the supersoft flux had decreased 
by 5 orders of magnitude comparing with the estimates of \citet{rau10V4743Sgr}, but the 
equivalent blackbody temperature had not decreased, implying shrinking of the emitting region.
In other cases, novae have been observed to become fainter at decreasing temperature, consistently
 with cooling as hydrogen burning turns off \cite[e.g. V1974 Cyg,][]{bal98V1974Cyg}. 

We wanted to assess whether freezing the value of the column density N(H) at a higher assumed
value results in a fit with a lower temperature and higher luminosity, indicating that the
whole WD surface may be emitting. The value of the interstellar absorption, derived in our
fit is consistent with the estimates of \citet{rau10V4743Sgr}, but is about two times lower than
the one in the direction of V4743 Sgr given by the Leiden/Argentine/Bonn (LAB) Survey
of Galactic H I and the Dickey\&Lockman H I map in the N(H) ftool\footnote{
http://heasarc.gsfc.nasa.gov/cgi-bin/Tools/w3nh/w3nh.pl} (a column density of 1.05 
and 1.41$\times10^{21}$\,cm$^{-2}$, respectively). Moreover, \citet{van12V4743Sgr} claimed 
that the analysis of the red tail slope of the {\it Chandra} X-ray spectrum of V4743 Sgr in
outburst, indicates N(H)$=1.36\pm0.04\times10^{21}$\,cm$^{-2}$. So there are indeed reasons to
think that N(H) may be higher than our best fitting values.
We thus set the value of N(H) to 1.36$\times10^{21}$\,cm$^{-2}$ and fitted the 2004 spectrum 
again with both the blackbody and the atmospheric model. As expected, the increased absorption
mainly affected the supersoft component, and returned T$_{\rm bb}=38.7_{-2.7}^{+3.1}$\,eV, 
T$_{\rm atm}=662_{-21}^{+20}$\,kK, and much higher luminosity --- 
L$_{\rm bb}=1.1_{-0.4}^{+0.7}\times10^{35}~{\rm D}^2$$_{\rm 4 kpc}$\,erg\,s$^{-1}$,
 L$_{\rm atm}=2.55\times10^{34}~{\rm D}^2$$_{\rm 4 kpc}$\,erg\,s$^{-1}$,
corresponding to the radius of the emitting region of
R$_{\rm bb}=6.2\times10^7~{\rm D}$$_{\rm 4 kpc}$\,cm and 
R$_{\rm atm}=1.3\times10^7~{\rm D}$$_{\rm 4 kpc}$\,cm, respectively, still smaller 
than the WD radius (for our calculations we assumed a distance of 4 kpc, since this value is close to the 
mean of the estimates mentioned in the Introduction).
Another mechanism that should be taken into account is a possible absorption in the 
accretion curtain of the magnetized WD. In our best fitting model we assumed that the partial
covering absorber affects only the thermal plasma emission, however, we will further show that
the 2004 soft X-ray light curve is also modulated with the WD spin period, which may be due
to the accretion curtain, crossing the line of sight (see fig. \ref{fig:softhard} and Section \ref{timing}). 
If we assume that the blackbody component is absorbed by the partially covering absorber, 
the resultant blackbody luminosity is 
L$_{\rm bb}=3_{-2}^{+10}\times10^{35}~{\rm D}^2$$_{\rm 4 kpc}$\,erg\,s$^{-1}$ and R$_{\rm bb}=1.8\times10^7~{\rm D}_{\rm 4 kpc}$\,cm.

\begin{figure}
\centering
\includegraphics[width=140pt, angle=-90]{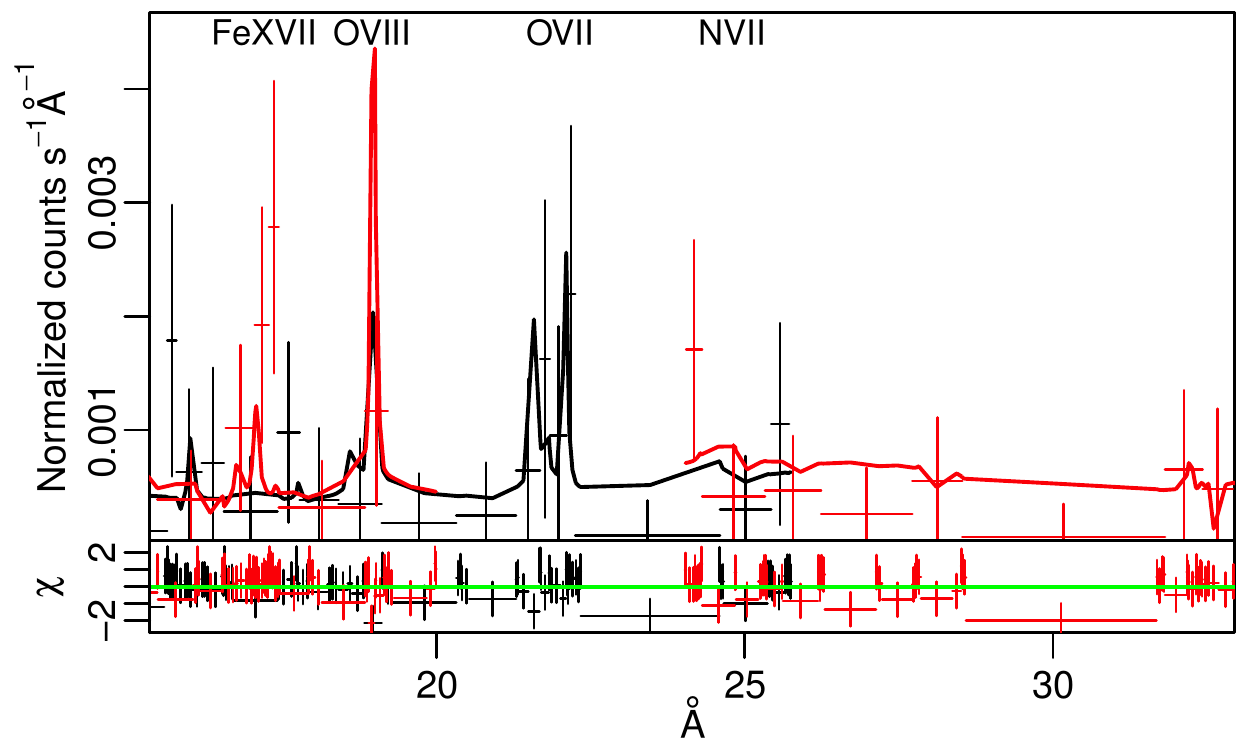}
\caption{ Top panel: The 2004 XMM {\sl Newton} RGS spectra of V4743 Sgr. 
The RGS 1 data are plotted in black and RGS 2 --- in red. The solid lines represent the 
best-fitting model of the XMM {\sl Newton} MOS data. The data were binned for visualizing purposes . Bottom panel:
residuals from the model.}
\label{fig:rgs}
\end{figure}

The 2004 RGS spectra are presented in fig \ref{fig:rgs}. Several emission lines are
marginally detected: O VII at $\sim$21.8--22 {\AA}, O VIII at $\sim$18.9 {\AA}, Fe XVII at
17.2 $\AA$ and N VII at $\sim$25 {\AA}.
The spectrum has a very low signal-to-noise ratio and we did not use it to refine the model
parameters. In fig. \ref{fig:rgs} we overplot the best-fitting model over the binned 
RGS 1 and 2 data and present the residuals in order to show that the model is consistent 
with the RGS data.
  
\subsection{Timing analysis}\label{timing}
\begin{figure}
\centering
\includegraphics[width=\columnwidth]{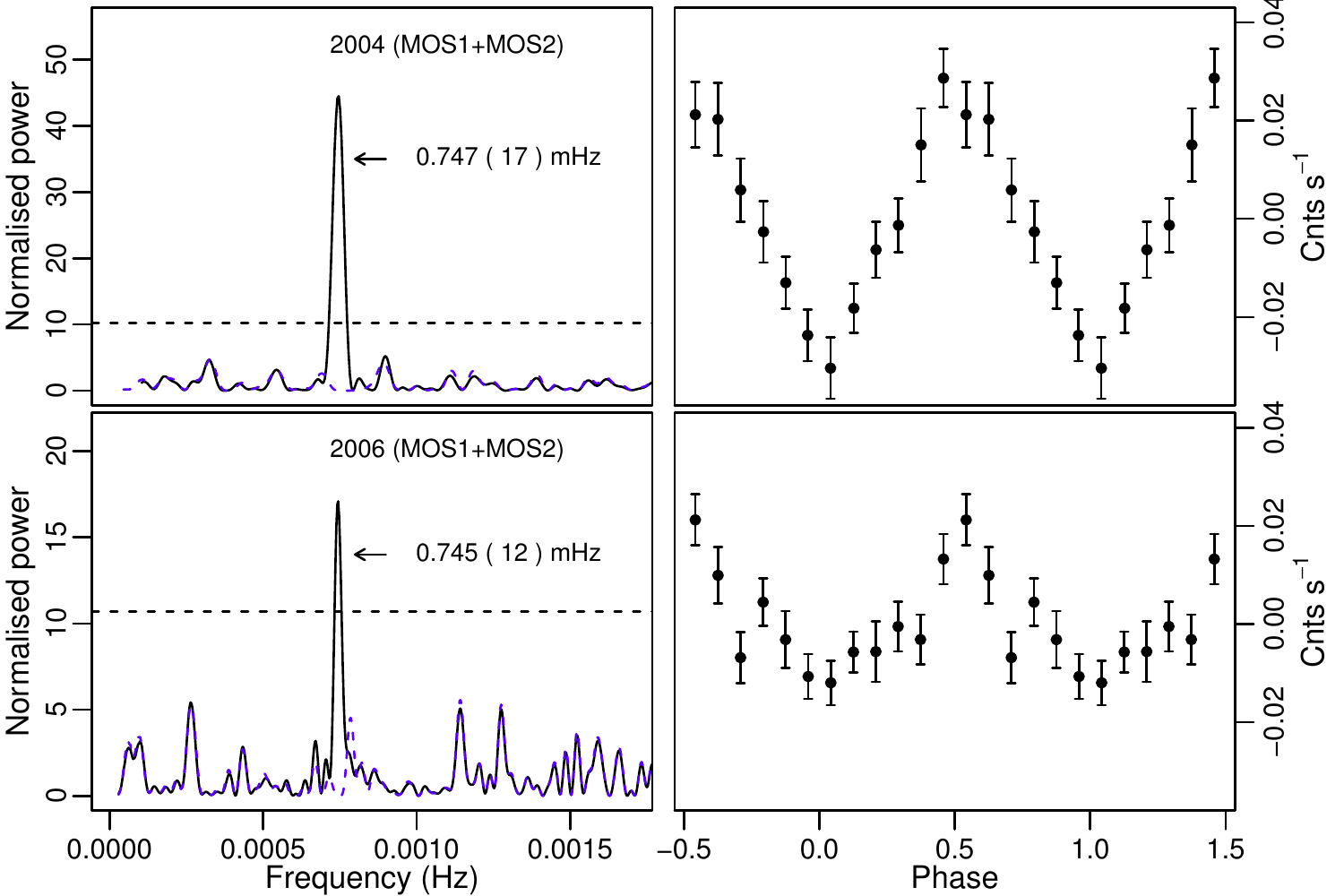}
\caption{The Lomb-Scargle periodograms and the phase folded light curves of the 2004 (top) and 
2006 (bottom) {\sl XMM Newton} light curves. The data
from the MOS 1 and MOS 2 detectors were combined. The values of the frequencies and the 
1$\sigma$ errors, calculated by fitting a Gaussian in the main peak of the LSP,
are marked in the plots. The horizontal dashed line represents the 
0.3\% false alarm probability level. The blue dashed lines are the LSPs of the same datasets 
after subtraction of the highest peak. The light curves were folded with the same frequency -- 0.748 mHz.}
\label{fig:lsp}
\end{figure}

Although \citet{dob10V4743Sgr} included the data we present here, we have performed our own, 
independent, timing analysis of the X-ray light curves in order to quantitatively confirm
the existence of the WD spin modulation using the bootstrap method and to study a possible 
energy dependence of the signal. We binned the {\sl XMM Newton} light curves every 100 
seconds, combined the data of the MOS 1 and MOS 2 detectors and subtracted a possible 
long-term trend with a 3-order polynomial fit. We applied the Lomb-Scargle method \citep{sca82LS}
for the 2004 and 2006 datasets in order to investigate the X-ray intensity modulations. 
The resultant Lomb-Scargle periodograms (LSPs) are presented in Figure \ref{fig:lsp}. 
In both light curves the $\sim$0.75 mHz frequency is clearly detected. The dashed horizontal line is the 
0.3\% false alarm probability and the dashed blue line shows the LSP after the subtraction
of the main peak by fitting and removing a sine function. We also applied the bootstrap
method, repeatedly scrambling the
data 10000 times and calculating the probability that random peaks in LSP in the range 0.1 mHz -- 2 mHz 
exceed the hight of the peak at 0.75 mHz in the original LSP. The probability that the peak
in the original data occurred by chance is zero for the 2004 dataset and 0.02\% 
for the 2006 one.
In order to {confirm the values of the frequencies with an independent analysis} we applied
the Phase Dispersion Minimisation (PDM) method \citep{PDM} to the same light curves. The values of the 
frequencies found with different methods are presented in Table 3.
To summarize, we confirm the broadband timing properties as published by \citet{dob10V4743Sgr}.

\begin{figure}
\centering
\includegraphics[width=\columnwidth]{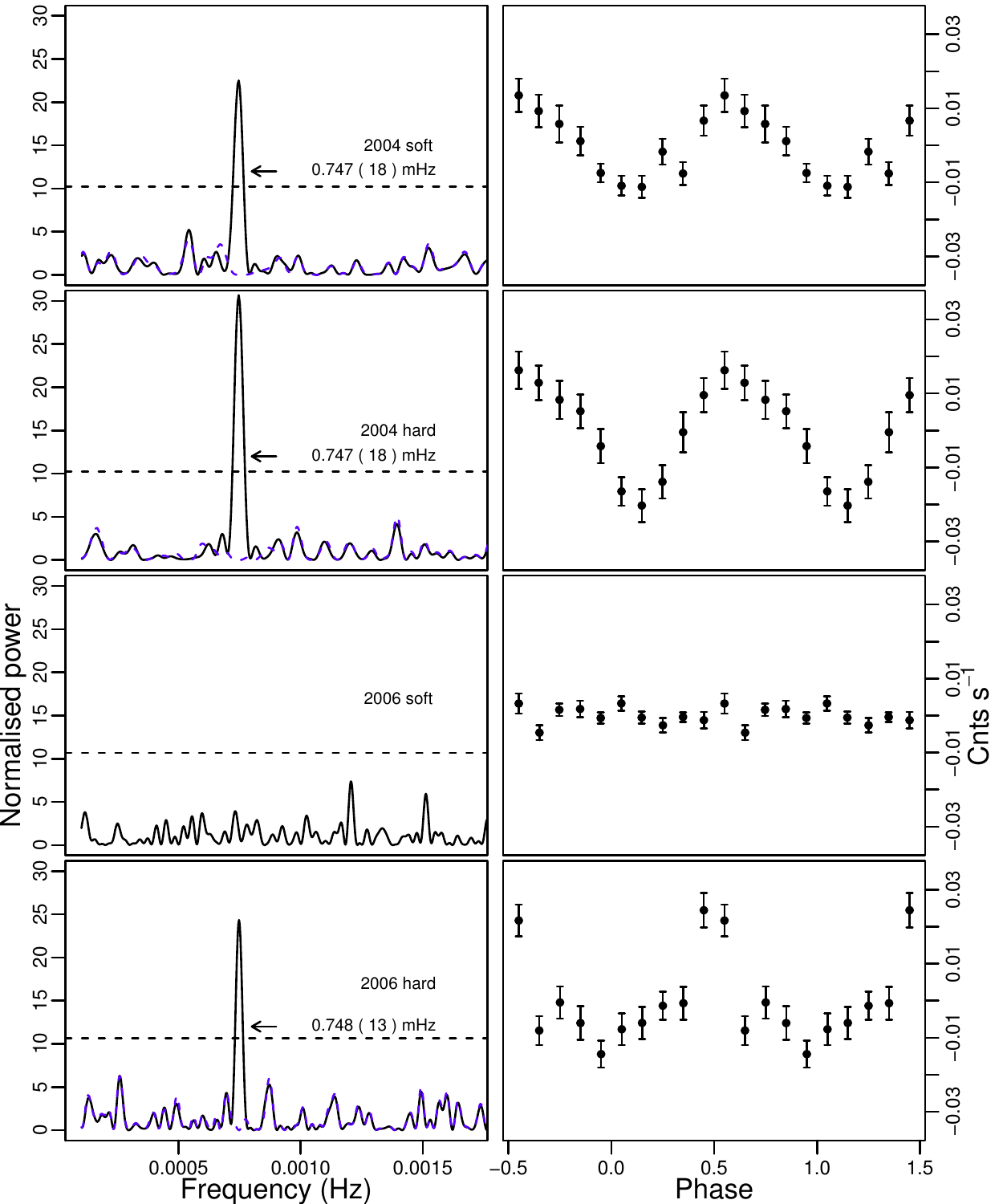}
\caption{ The same as fig. \ref{fig:lsp}, but for energy ranges below (soft) and above 
(hard) 0.8 keV.}
\label{fig:softhard}
\end{figure}

We investigated a possible energy dependence of the X-ray modulation. We first extracted 
the light curves in two energy ranges: 0.3 -- 0.8 keV and 0.8 -- 10 keV and performed 
the same methodology as was mentioned above. Fig. \ref{fig:softhard} represents the 
LSPs for the hard and soft energy ranges of the 2004 and 2006 datasets. While in 2004 
the modulation was the same in both energy ranges, in 2006 it was present only in the hard
X-rays. The amplitudes of the modulation in 2004 were $\sim$50\% and 30\% in the soft an hard ranges,
respectively and $\sim$30\% in the hard range in 2006.
Aiming to constrain the spectral component, modulated with the spin period in the 2004 dataset, we extracted the 
MOS 1 and 2 light curves in a narrower range --- 0.3 -- 0.6 keV, where the blackbody-like emission
dominates. The modulation was still present even in this energy range, indicating that 
the supersoft X-ray emission is modulated with the same period.

Although the orbital modulation was observed in the optical band \citep{kan06V4743Sgr,ric05V4743Sgr, wag03V4743Sgr}, 
indicating a moderately high inclination of $\sim60^\circ$, we find no variability that might represent an orbital 
modulation in our X-ray data, neither in the soft, nor in the hard ranges.
 
\section{SALT observations}\label{ospec}
Optical spectra of V4743 Sgr in the $\lambda\lambda$4500--5600 {\AA} range were obtained 
on March 21 2014 at the SALT telescope using the Robert Stobie Spectrograph (RSS), grating PG2300, in a long slit mode (single, 8 arcmin long,
 1.5 arcsec wide slit)\footnote{
Under program 1178-7 2013-2-UW-001 (PI: Marina Orio)}. The instrumental resolution is 110 -- 120\,km\,s$^{-1}$.
For the flux calibration we used the standard star Feige 110. 
This observation was done in the framework of a monitoring program of novae previously observed as SSS
as they returned to quiescence (Zemko 2016 in prep). 

\begin{figure}
\centering
\includegraphics[width=\columnwidth]{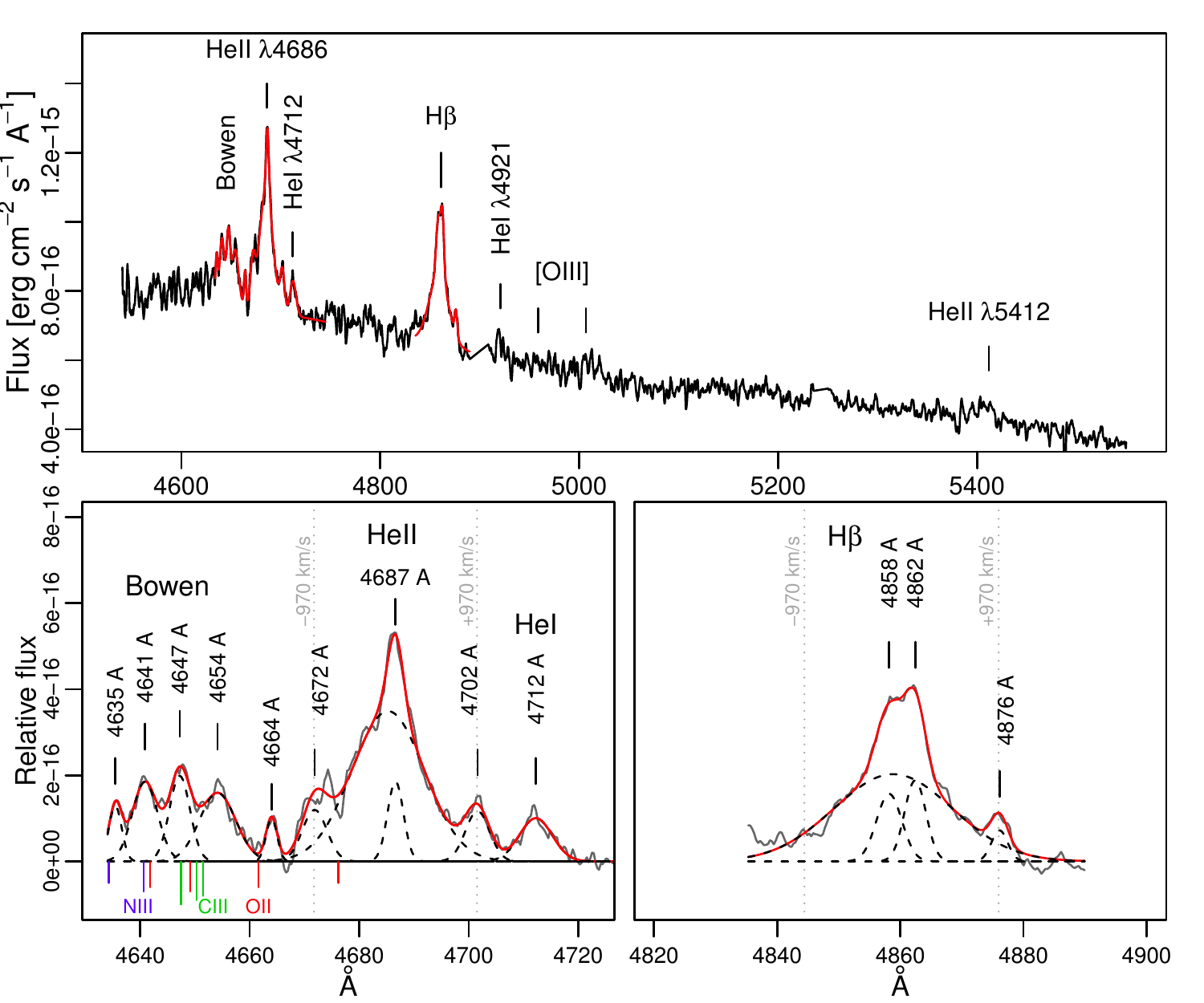}
\caption{Top: The SALT spectrum of V4743 Sgr, revealing the strong emission lines of the He II
$\lambda$4686 and H$\beta$. The Bowen blend $\lambda\lambda$4640 -- 4650 is a combination 
of high excitation lines, mainly C III, O II, N III. Bottom: The regions of the 
 Bowen blend, He II $\lambda$4686 and the H$\beta$ lines
 after the subtraction of the continuum.
The red lines show the fit of the He II $\lambda$4686, Bowen blend and He I 
(left) and the H$\beta$ (right) lines. Both the He II $\lambda$4686 and H$\beta$ lines
have a narrow and a broad component.
The central positions of the best fitting Gaussians are marked on the plots (excepting the 
broad components, since they roughly coincide with the narrow ones). The grey dotted vertical 
lines show the velocity shifts of $\pm$970 km s$^{-1}$. The black dashed lines 
represent the Gaussians that were introduced to the fit. The table positions of
 the N III, O II and C III lines that constitute the Bowen blend are marked
with blue, red and green vertical lines, respectively. The length of the label depends on the 
 laboratory intensity of the emission line.}
\label{fig:salt}
\end{figure}

The flux calibrated and dereddened optical SALT spectrum is
presented in the top panel of Figure \ref{fig:salt}. For dereddening we assumed 
N(H)=0.71$\times10^{21}$\,cm$^{-2}$ (see Table 2), which corresponds
to E({\it B}-{\it V})=0.12 \citep{boh78abs}.
The [O III] $\lambda$5007 line is only marginally detected (the rest wavelengths of the 
[O III] lines are marked on the plot), which is typical of a very fast nova after 10 years 
\citep{dow01shells}. 

The strongest emission lines are He II $\lambda$4686, H$\beta$ and the Bowen blend. 
We removed the continuum using the {\it continuum} task in IRAF and fitted with Gaussians 
the main features in these regions (the H$\beta$, He II
$\lambda$4686 lines and their components and the Bowen blend) in order to measure their central
 positions and velocity broadening. 
The result of the fit is shown with the red line in all the plots of Figure \ref{fig:salt} 
and the dashed black lines on the bottom plots show the Gaussian components that were introduced. 
The central positions of the best fitting Gaussians are also marked.
From this fit we found that both H$\beta$ and He II $\lambda$4686 emission lines have
a narrow and a broad component, just slightly shifted with respect to each other as it is seen
from the bottom panels of Fig. \ref{fig:salt}. The narrow component of H$\beta$ has a double
peaked profile with a separation of only $\sim$250\,km\,s$^{-1}$ between the two peaks,
 which is quite small for the accretion disk rotation 
(assuming M$_{\rm WD}$=1.2 M$_{\odot}$, P$_{\rm orb}$=6 h and a reasonable value for the 
mass ratio q$\sim$0.8 the velocity will be about 560\,km\,s$^{-1}$ at the radius of the 
tidal limit of the accretion disk, which is expected to be an upper limit for the disk radius,
see e.g. \citealt{war95book}). The broad component of H$\beta$ has a FWHM of $\sim$1300\,km\,s$^{-1}$ 
and the broad component of 
He II $\lambda$4686 --- $\sim$990\,km\,s$^{-1}$. The H$\beta$ line seems to have another small
component at 4876 {\AA}, red-shifted by $\sim$970\,km\,s$^{-1}$ with respect to the
position of the double-peaked central line. Interestingly, similar small components, red and blue-shifted
by almost the same velocity of $\sim$950\,km\,s$^{-1}$ are also observed in the He II $\lambda$4686 line.

Another prominent feature of the spectrum is the Bowen blend at $\lambda\lambda$4640--4650 {\AA}, 
which is a combination of three lines of C III, two of N III and eight of O II 
\citep{mck75bowen}. The rest positions of the most intensive lines are indicated in the 
bottom-left panel of Figure \ref{fig:salt}. 
The length of the label of each line depends on its laboratory intensity: the higher the intensity is 
the bigger is the label. We see that the $\lambda$4641 and $\lambda$4647 lines, observed in the 
spectrum, coincide with the rest positions of the N III and C III lines, respectively.
On the other hand, the origin of the $\lambda$4654 line is unknown.
The equivalent widths (EWs) of the  He II $\lambda$4686 and H$\beta$ emission lines 
are about 10.5 $\AA$ and that of the Bowen blend is $\sim$5.2 $\AA$.

\section{Discussion}{\label{seq:disc}}
V4743 Sgr is a bright X-ray source, which emitted both soft ($<$0.6 keV) and hard X-rays in 2004
and only hard in 2006. Novae shortly after eruption can emit X-rays in the 1--10 keV range
{originating from the shocked ejected shells \citep{muk08Novae, nes13SSS},
but since the source of this emission is spatially extended it cannot be variable on short time scales,
in contrast to our observations (fig. \ref{fig:softhard} and \ref{fig:lsp}). Basing on the fact that
the hard X-ray component, which is present in both observations, is modulated with the WD spin period
and is also well fitted with the two-temperature thermal plasma emission model, typical of
plasma in collisional equilibrium cooling as it settles onto the WD, we argue that this component 
is due to the resumed accretion.
 
The disappearing of the soft emission in 2006 indicates that it is, in turn, associated
with hydrogen burning and not with accretion. The soft component was also modulated
with the WD spin period supporting the idea that the source of this emission is close to the WD surface, 
like in case of short-period X-ray oscillations of novae in SSS phases found by \citet{nes15Oscil}. 
Since both {\sl XMM Newton} exposures are longer than the orbital period the disappearing 
of the supersoft component is also not an effect of different orbital phases.

The He II $\lambda$4686 line and the Bowen blend in the {\sl SALT} spectra show that there is still an 
ionizing component that is strong in the 200--300$\AA$ range. This component may be the same 
that was previously emitting the supersoft X-rays and has cooled to a peak temperature in the UV range.
If so, the disappearance of the supersoft X-ray emission in 2006 is probably consistent with cooling.
The EWs of the He II $\lambda$4686 and H$\beta$ lines are small in comparison to what is usually
observed in novae a decade after explosions \citep{rin96novaspec, tom15SALT}. It may imply
that the WD photosphere is cooling fast, reducing the number of photons with wavelengths 
shorter than the 228$\AA$ edge, consequently contributing less to the formation of the 
He II $\lambda$4686 line. Alternatively, the small EWs may be a result of a strong reprocessed continuum
due to irradiation of the accretion disk and the secondary.

In this section we will discuss possible emitting sites of the soft X-ray component and 
magnetic nature of V4743 Sgr.


\subsection{The IP scenario}
Several observational facts suggest that V4743 Sgr may be an intermediate polar.
\begin{itemize}

\item Taking into account uncertainty in the distance, discussed above, the hard X-ray 
luminosity in the 2.0--10 keV range was $2.4\times10^{32}$--$6.2\times10^{33}$\,erg\,s$^{-1}$
in the 2006 dataset, which is typical of IPs \citep{pre14IP}.

\item The spectra can be fitted only introducing a complex, partially covering absorber, 
which can be a result of periodical obscuration of the central emitting region by accretion curtains
\citep{eva07softIP, anz08softIP}. 

\item The most important indication of a magnetic nature of V4743 Sgr is the presence of 
the coherent $\sim$0.75 mHz modulation, observed in X-rays even in quiescence. The frequency
was stable from 2004 to 2006, to within measurement limitations. The WD rotation
in IPs is not synchronized with the orbital period and once during the spin period the accretion
curtain crosses a line of sight, resulting in the X-ray light curve pulsations. 
Since the X-rays are also reprocessed from the surface of the secondary, the beat period between
the spin and the orbital one is often observed in the optical band. 
Taking into account two periodicities found in the optical
observations \citep{kan06V4743Sgr} and the WD spin period measured in our X-ray data, we see that a relation 
$1/{\rm P_{\rm spin}}-1/{\rm P_{\rm orb}}=1/{\rm P_{\rm beat}}$ is perfectly satisfied.
However, the orbital period is yet to be spectroscopically confirmed.

\item The prominent He II $\lambda$4686 line and the high value of $EW$(He II $\lambda$4686)$/EW$(H$\beta$)
ratio are typical of IPs \citep[the ratio is $\sim$1, while in IPs it is usually $>$0.4,][]{sil92IPHe2}, 
although this is not observed {\it only} in IPs\footnote{ for a
discussion see \\
http://asd.gsfc.nasa.gov/Koji.Mukai/iphome/issues/heii.html}. Most 
interestingly, the emission lines have a complex structure with red and blue-shifted components with the same velocity, 
which may originate in the accretion curtains in the magnetosphere of the WD as it was
observed in another IP and an old nova -- GK Per \cite{bia03GKPer}. 
Since the total exposure time of two SALT spectra of V4743 Sgr (which were combined for the analysis) covers 
more than one spin period all the possible WD spin-depended effects were smeared out
in the resultant spectrum. SALT does not usually allow phase-constrained observations, but
in the future we would like to obtain phase resolved optical spectra with another telescope,
to better constrain the emission site of these lines components.

\end{itemize}

The only element that seems odd and somehow out of place in the IP scenario is that the soft
 X-ray modulation disappeared in the 2006 data (see fig. \ref{fig:softhard}): the variability
was observed only above 0.8 keV. This is in contrast with the usual situation for IPs in 
which X-ray modulations are more prominent in soft X-rays, since the cross section of 
photoelectric absorption in the accretion curtain decreases with energy. 
We also did not detect any modulation of the X-rays related to the orbital period, but this
can be due to relatively short exposure times. 
In fact, during the outburst, there was
a marginal detection of orbital modulation in the supersoft X-rays light curve \citep{lei06V474Sgr}.
The orbital modulation in the X-ray light curve is usually, but not always, observed in IPs
 with inclination angles exceeding $\sim60^\circ$ \citep{par05IPorb}.
Altogether, many observational facts indicate that V4743 Sgr is an IP and we favour this hypothesis.

\subsection{The soft X-ray component}

Taking into account all the measurements of T$_{eff}$ and the luminosity of the supersoft X-ray
component in the 2004 {\sl XMM Newton} spectrum (see Tab. 2 and the section \ref{sseq:xrayspec}) we find
that the luminosity decreased at least by two orders of magnitude compared with the 
values found by \citet{van12V4743Sgr} and \citet{rau10V4743Sgr} in the last Chandra exposure
during the SSS phase and the estimates of the emitting region radius lie in the range 
$4.0\times10^6 - 6.2\times10^7~{\rm D}$$_{\rm 4 kpc}$\,cm.
Even the largest possible distance of 6 kpc (found by \citealt{nie03v4743sgr}) would
 only increase this value by 50\%.
It is not consistent with an emitting region as large as the whole WD surface but is close
to the estimates of the size of a polar cap in an IP \citep{hel97IPcap}.

A similar low luminosity blackbody component, with a temperature of about 70 eV, was detected
in V2491 Cyg more than 2 years after the outburst \citep{pag10V2491cyg, zem15V2491Cyg}. 
The size of its emitting region decreased well below the size of the surface of the WD. 
Since V2491 Cyg was also an IP
candidate, \citet{zem15V2491Cyg} proposed that the blackbody emission can originate either 
in the irradiated polar cap, like in ``soft IPs'' or, less probable, from a localized residual 
hydrogen burning, left after the nova explosion. Another example is V2487 Oph --
a recurrent nova and an IP candidate -- which showed blackbody-like emission in X-rays even 
8.8 years after the nova explosion \citep{her14V2487Oph}. However, in the 2006 observations of V4743 Sgr
the blackbody component was not detected and it would be difficult to explain this fact in 
 the context of the soft IP scenario. Irradiation of the polar cap
region can indeed depend on the mass transfer rate, but since the hard X-ray component was the
 same in both datasets -- 2004 and 2006 -- we do not expect the mass transfer rate to vary significantly.
A possible explanation is a temperature gradient on the surface of the WD, which 
can remain for several years after the nova explosion. It is not clear, however, why in V4743 Sgr
this gradient disappear already after 3.5 years, while in V2487 Oph it was observed for almost
9 years.

\section{Conclusions}
We analysed two quiescent {\sl XMM Newton} observations of V4743 Sgr in the first years after 
the outburst, and a later optical spectrum. A stable, coherent modulation of the X-ray light curves at frequency
$\simeq$0.75 mHz, observed in both {\sl XMM Newton} exposures, and the X-ray spectra indicate that V4743 Sgr is 
an IP. The optical spectrum revealed prominent emission lines of He II $\lambda$4686, H$\beta$ 
and the Bowen blend. The EWs of the emission lines are low, implying that the WD photosphere
 was cooling relatively fast or that there is a strong reprocessed continuum
due to irradiation of the accretion disk and the secondary. The He II $\lambda$4686 line 
and the Bowen blend point towards a hot region, which 
may be the same that was previously emitting the supersoft X-rays and has simply cooled
to a peak temperature in the UV range. We found that the He II $\lambda$4686 and H$\beta$ 
emission lines of V4743 Sgr have a complex structure with a narrow, broad and blue and 
red-shifted satellite components, which may be due to the ``accretion curtains'' in an IP.
The $EW$(He II $\lambda$4686)$/EW$(H$\beta$) ratio is high, which is consistent with the IP 
interpretation.

In the X-ray spectrum obtained 2 years after the explosion we detected a supersoft component at
a temperature close to the WD temperature during the SSS phase, but with at least two orders of 
magnitude lower luminosity. In contrast to the typical behaviour of novae, V4743 Sgr seems
to show a shrinking of the soft X-rays emitting region at almost constant temperature instead
 of cooling. So far, we know 
only two such examples: V2491 Cyg and V2487 Oph that are both IP candidates and most probably
host massive WDs \citep{her14V2487Oph, zem15V2491Cyg}. In V4743 Sgr the supersoft X-ray component 
disappeared by the time of the last {\sl XMM Newton} 
observation, implying that the source of radiation is not related to accretion and cannot
be explained by the irradiated polar cap on the WD, like in ``soft IPs''. We propose that 
this supersoft emission may be due to a temperature gradient on the surface of the WD, which 
can be a characteristic of magnetic novae.

The disappearing low-luminosity supersoft component is a very interesting phenomenon that 
we still do not quite know how to explain. We know that such a component was observed in two
other novae, and that in V2487 Oph it lasted for at least almost 9 years. Data on other 
novae are still missing and further X-ray monitoring is encouraged, because this phenomenon
may have profound implications for the secular history of accretion and hydrogen burning.

\begin{table*}
\centering
\begin{minipage}{140mm}
  \caption{The parameters of the best fitting models -- wabs$\times$(bb/atm+pcf$\times$(vapec+vapec)) for V4743 Sgr X-ray spectra.
 The errors represent the 90\% confidence region for a single parameter. The luminosity is given assuming a distance of 4 kpc, consistent with the MMRD relation (see discussion in the text).}
\label{tab:fit}
	\begin{tabular}{lllllll}
	\hline
								 							 &\multicolumn{4}{c}{Simultaneous fits}			  							& \multicolumn{2}{c}{Independent fits} \\
	\hline
Parameter						 & 2004					 	& 2006  				 & 2004+atm					& 2006+atm  			& 2004				  	& 2004+atm.			 	\\
	\hline
N(H)$^a$						 &$0.71_{-0.25}^{+0.33}$ 	&$0.71_{-0.25}^{+0.33}$  &$0.40_{-0.16}^{+0.18}$ 	&$0.40_{-0.16}^{+0.18}$ &$0.4_{-0.3}^{+0.4}$ 	&$0.24_{-0.17}^{+0.18}$ \\
N(H)$_{\rm{pc}}$$^b$ 			 &$87_{-17}^{+20}$	 	 	&$57_{-28}^{+55}$		 &$81_{-16}^{+20}$	 	 	&$40_{-16}^{+38}$		&$85_{-17}^{+20}$  	 	&$80_{-16}^{+19}$		\\
CvrFract$^c$ 					 &$0.75_{-0.05}^{+0.03}$ 	&$0.46_{-0.08}^{+0.14}$  &$0.75_{-0.05}^{+0.03}$ 	&$0.45_{-0.06}^{+0.12}$ &$0.75_{-0.05}^{+0.03}$	&$0.75_{-0.03}^{+0.03}$ \\
T$_{\rm bb/atm}$ (eV)			 &$53_{-9}^{+10}$		 	& 	 					 &$63_{-3}^{+4}$		  	& 						&$64_{-15}^{+15}$	 	&$67_{-4}^{+6}$	 	 	\\
T$_1$ (keV)						 &$0.206_{-0.020}^{+0.027}$	&$0.105_{-0.018}^{+0.040}$&$0.202_{-0.018}^{+0.023}$&$0.13_{-0.03}^{+0.03}$	&$0.21_{-0.03}^{+0.04}$	&$0.217_{-0.023}^{+0.028}$\\
T$_2$ (keV)						 &$13_{-3}^{+12}$		  	&$>$14	  	 			 &$14_{-4}^{+12}$		  	&$>$17	  	 			&$13_{-3}^{+12}$		&$13.3_{-3.0}^{+13}$	\\
Norm$_1$$^d$					 &$0.33_{-0.16}^{+0.26}$	&$0.40_{-0.29}^{+1.13}$  &$0.30_{-0.11}^{+0.17}$	&$0.10_{-0.03}^{+0.35}$ &$0.18_{-0.14}^{+0.26}$	&$0.21_{-0.09}^{+0.14}$ \\
Norm$_2$$^d$					 &$1.49_{-0.14}^{+0.16}$ 	&$1.02_{-0.16}^{+0.17}$  &$1.44_{-0.14}^{+0.16}$ 	&$1.03_{-0.18}^{+0.14}$ &$1.49_{-0.14}^{+0.16}$	&$1.44_{-0.14}^{+0.16}$ \\
Flux$_{\rm abs}^e$ 			 	 &$2.28_{-0.32}^{+0.07}$ 	&$1.56_{-1.5}^{+0.5}$ 	 &$2.29_{-2.0}^{+0.05}$ 	&$1.6_{-1.5}^{+1.2}$ 	&$2.29_{-0.3}^{+0.08}$ 	&$2.3_{-1.9}^{+4.5}$ 	\\
Flux$_{\rm unabs}^e$ 			 &7.62 					 	&2.88 				 	 &5.35 						&2.13 				 	&5.35 				 	&4.82				  \\
Flux$_{\rm abs}^{\rm bb/atm}$$^f$&0.282  				  	& 	 					 &0.227  				  	& 	 					&0.306  				&0.245				 	\\
Flux$_{\rm unabs}^{\rm bb/atm}$$^f$&1.98	  				& 						 &0.552	  					& 	 					&0.981	  			 	&0.431				 	\\
$\chi^2$						 &1.2	 	  			  	&1.2 					 &1.2	 	  			  	&1.2 					&0.9	 	  			&0.9 					\\ 
L$_{\rm 2.0-10.0 keV}^g$ 		 &4.62					 	&2.76				 	 &4.53					 	&2.70				 	&4.60				  	&4.52				 \\
L$_{\rm bb}$$^h$ 				 &$8_{-5}^{+25}$	  		& 						 &4.9	  					& 						&$3.2_{-1.6}^{+12.8}$ 	&3.84					\\
R$_{\rm bb/atm}$$^i$ 			 &8.8					  	& 						 &$5_{-3}^{+4}$				& 						&3.8					&$3.9_{-2.7}^{+3.4}$	\\
\hline 
\multicolumn{7}{p{.9\textwidth}}{$^a$ $\times10^{21}$\,cm$^{-2}$. }\\
\multicolumn{7}{p{.9\textwidth}}{$^b$ N(H)$_{\rm{pc}}$ $\times10^{21}$\,cm$^{-2}$ for the partial covering absorber.}\\
\multicolumn{7}{p{.9\textwidth}}{$^c$ Covering fraction of the partial covering absorber.}\\
\multicolumn{7}{p{.9\textwidth}}{$^d$ $\times10^{-3}$ Normalisation constant of the VAPEC model.}\\
\multicolumn{7}{p{.9\textwidth}}{$^e$ The X-ray flux ($\times10^{-12}$ erg cm$^{-2}$ s$^{-1}$) measured in the range 0.2--10.0 keV. The Flux$_{\rm unabs}$ represents the value
of the X-ray flux, corrected for the interstellar and intrinsic absorption.}\\
\multicolumn{7}{p{.9\textwidth}}{$^f$ The X-ray flux ($\times10^{-12}$erg cm$^{-2}$ s$^{-1}$) of the 
blackbody component measured in the range 0.2--10.0 keV.}\\
\multicolumn{7}{p{.9\textwidth}}{$^g$  The X-ray luminosity ($\times10^{33}$\,erg\,s$^{-1}~{\rm D}^2$$_{\rm 4 kpc}$)
in the range 2.0--10.0 keV. The L$_{\rm 2.0-10.0 keV}$ was calculated from the X-ray flux 
in the range 2.0--10.0 keV, corrected for the interstellar and intrinsic absorption.}\\
\multicolumn{7}{p{.9\textwidth}}{$^h$  The bolometric X-ray luminosity ($\times10^{33}$\,erg\,s$^{-1}~{\rm D}^2$$_{\rm 4 kpc}$)
of the blackbody and atmospheric component. L$_{\rm bb/atm}$ were calculated from the normalization constants of the models.}\\
\multicolumn{7}{p{.9\textwidth}}{$^i$  The radius of the emitting region ($\times10^{6}$\,cm). For the blackbody fit R$_{\rm bb}$ was found from the
Stefan-Boltzmann law. The atmospheric model gives the value of the emitting radius $R_{\rm atm}=10^{-11}\sqrt{\rm norm}~{\rm D}_{\rm 4 kpc}$ cm.}\\
\hline 
\end{tabular}
\end{minipage}
\end{table*}

\begin{table}
\centering
\begin{minipage}{80mm}
  \caption{The results of the timing analysis of the {\sl XMM Newton} light curves.}
  \begin{tabular}{lcc}
  \hline
Dataset							& \multicolumn{2}{c}{Frequency (mHz)} \\
								& LS		& PDM 		 \\
\hline
2004 (MOS1+MOS2) 				& 0.747(17)	& 0.7424(14) \\
2006 (MOS1+MOS2)				& 0.745(12)	& 0.7434(28) \\
\hline 
\multicolumn{3}{p{.9\textwidth}}{{\bf Notes} LS --- Lomb-Scargle method; PDM --- Phase Dispersion
Minimisation method. The erros in the frequencies found with the LS 
method correspond to the 1$\sigma$ level.}\\
\hline 
\end{tabular}
\label{tab:freq}
\end{minipage}
\end{table}

\section*{Acknowledgements}
Some of the observations reported in this paper were obtained with the Southern African
Large Telescope (SALT). Polina Zemko acknowledges a pre-doctoral grant of the CARIPARO 
foundation at the University of Padova. Dr. Orio was funded by the NASA {\sl XMM Newton}  program. 


\bsp
\label{lastpage}
\end{document}